# Low-Cost Bluetooth Mobile Positioning for Location-based Application

Zaafir barahim
Computer Science and
Engineering Department
University of Mauritius
Reduit, Mauritius
b.zaafir@gmail.com,

M. Razvi Doomun
Computer Science and
Engineering Department
University of Mauritius
Reduit, Mauritius
r.doomun@uom.ac.mu

Nazrana Joomun
Computer Science and
Engineering Department
University of Mauritius
Reduit, Mauritius
jn144@hotmail.com

*Abstract*— **Bluetooth is a promising short-range radio network technology. We present a low cost and easily deployed, scalable infrastructure for indoor location-based computing of mobile devices based on Bluetooth technology. The system consists of 2 main components, namely the Bluetooth (BT) Sensor System and the Central Navigation System which have been developed using the JDK 6.0. The Bluetooth Sensor System allows mobile devices whose Bluetooth mode is set to discoverable, to be scanned and detected, and they receive customizable text message of their positioning information, e.g. room identity. The positioning information is also sent to the Central Navigation System which in turn displays and updates the navigation map. The system is also used to track the movement of different BT mobile devices within the implemented environment.**

## I. Introduction

Location-Based Services (LBS) is expanding rapidly with the capability of mobile networks to determine locations of mobile devices accurately. LBS applications provide location, navigation, information, targeted advertising, notification and other services, where the awareness of the user is decisive. These applications are generally developed over positioning systems and the infrastructure is extended to support various task-specific location-dependent responses. There are many location determination methods and technologies that can be used; however no single method will operate effectively in all intended environments with sufficient accuracy [1]. Hence, a limitation for many existing systems is that they are tailored to support only one type of positioning technology like GPS, GSM, Bluetooth, IR or WLAN positioning. Other issues are specific of positioning sensing technologies: for example, outdoor versus indoor positioning, accuracy, energy usage, coverage, speed, availability, cost, privacy aspects, and so forth. Given all these parameters to control, no single location-sensing technology can become dominant. The choice of which location-sensing technology to employ depends on the type of usage of location information, and multiple technologies may coexist.

Indoor location systems and applications are still a research issue today. This paper presents the design of a low cost University of Mauritius mobile guide system (UoMGuide) for location positioning on the University of Mauritius (UoM) campus. In Section II, a brief review of related work on existing mobile guide systems and positioning techniques are given. In Section III, we summarize the main mobile guide system requirements and methodology. The system design and implementation modules are discussed in section IV. Finally, the discussion and conclusion are presented in section V and VI respectively.

## II. Related Work

One of the popular location-based systems is Place Lab [2] that works by listening for the transmissions of wireless networking sources like 802.11 APs, fixed BT devices, and GSM cell towers. They all use protocols which give beacons a unique or semi-unique ID. The beacon database provides this beacon location information to client devices. The Place Lab clients use live radio observations and cached beacon locations to form an estimate of their location.

For indoor environments, different technologies and techniques exist to provide location information in ad hoc networks which are cost effective and give good location results [3]. Trilateration technique [4] applies geometric properties of access points' location to compute mobile position. It uses Received Signal Strength Indication (RSSI), propagation time and angulation to calculate location. Techniques for locating Wi-Fi devices include RSSI, time of arrival, angle of arrival, time difference of arrival (TDOA), and other radio-location methods [4]. Although Wi-Fi has been used in [5] as the indoor positioning technology, it is unsuitable for indoor environments because it is influenced by geometry of location, multipath errors and non-line of sight propagation. Conversely, Bluetooth positioning is mainly achievable in indoor environments where many known location stationary devices can reside and inquiry or RSSI methods could be used. Current scenarios of Bluetooth-assisted LBS are based on loose location awareness. Ekahau positioning engine [6] is a wireless positioning server of location accuracy up to 1 m average. It is based on signal strength site calibration.

Strongly inspired by the Universal Location Framework [7] we use a layered structure for our system design, to make the system less complex and to reduce the interaction between separate layers. Location Computing System (LCS) is normally partitioned into Physical/Sensor layer, Data/Measurement layer, Integration/Fusion layer, Security/Privacy layer and Presentation/Application access layer. The key reason for the multilayer framework is to decouple the application from the evolution of future location technologies and to represent standard forms of location information for higher interoperability and compatibility. At physical/sensor level, position is captured by hardware equipment, e.g. by a terminal or sensors observing signal strengths or Cell Ids, respectively. Data level, this information is processed and new data is computed, e.g. the absolute location of an object in x, y coordinates.

III. SYSTEM METHODOLOGY

The location server is responsible for storing and managing user locations. The system is designed to work for indoor environments, as illustrated in figure 1. The signal strengths received from the Bluetooth node servers, during the location determination phase, are gathered as vectors samples. They are compared to the location-map and the "best" match is returned as the estimated user location. Location positioning for UoMGuide system characteristics are.

**Layered and Modular deployment:** The system architecture can combine several context-provider components, making them interoperate. Hence, the overall infrastructure ensures system interoperability and context sharing. By adding or removing context-provider components, each device could be adapted to specific application requirements.

**User mobility support:** The infrastructure need to provide an event service that can be utilized to detect location-changes.

**Maintainability and updates:** Environment profiles can be dynamically replaced or added according to application requirements. For instance, if a more efficient algorithm becomes available, the component is easily updated and plugged into the service infrastructure. Sensor independence minimizes the impact of new technologies on WiDat application.

**Extensibility and Scalability:** Modularity also makes the system flexible and easy to extend through the addition of new sensors, new devices, and new services. The layered scalable architecture is crucial due to context dynamicity and in particular of user profiles. Users prefer to customize their preferences at run time.

**Collaboration and context-sharing:** The infrastructure enables collaboration and context sharing among context provider components and context-aware applications. The major benefits are that device capabilities can be significantly enriched by additional context-sources, and the context representation can be enhanced. Furthermore, single devices can reduce processing and storage efforts, and they can better cope with context uncertainty by exploiting additional context data retrieved from neighbouring devices. Sharing of context information must be in compliance with the user permission profile.

Two Java Specification Requests [8], namely JSR-179 and JSR-82 support the deployment of location-aware applications for MIDP-compliant devices. MIDP is the Mobile Information Device Profile of the Java 2 Micro Edition (J2ME) platform that is nowadays supported by many mobile devices, such as smart-phones and PDAs. The specification defines functionalities to request and get a location result. The **LocationProvider** class represents a module that is able to determine the location of the device. This may be implemented by using any possible location methods, such as satellite based methods like GPS, cellular network based methods, short-range positioning methods like Bluetooth Local Positioning, etc. The application can specify criteria for selecting the location provider that better fulfil these criteria. By using the **LocationProvider**, the application can get location objects representing the location of the device at the time of the measurement. The location is represented by the Location object that contains the geographical coordinates (latitude, longitude, and altitude) and information about their accuracy, a timestamp and possibly information about speed and course of the terminal. For some location methods, the Location object may also contain textual address information, e.g. a building name or address. The package also includes a database of landmarks, meaning known physical locations. Indeed, the user can store commonly used locations in the database, such as users home, office, etc. Additionally, the JSR-82 aims to define a standard set of APIs that will enable an open software development environment for Bluetooth-based devices. The main functionality provided by this specification includes discovery, communication, and device management. As for the discovery API, it is used to register services, as well as to discover devices and services. The communication API is used to establish connections between devices. The device management API allows management and control of the communication connections. The main advantages of such an approach is that these specifications could be implemented by means of existing location methods, including satellite based methods like GPS, as well as short-range positioning methods. For example in [1][6], the authors show how to extend the Java APIs for Bluetooth (JSR-82) in order to provide the Location API with RSS (Received Signal Strength)-based position-information. A new component, namely the

RSSI_Provider, is inserted into the JSR-82 API. It provides location information needed to build the Location objects. Thereby, a LocationProvider class (as defined in JSR-179) uses the RSSI_Provider to generate Locations. The main advantage of such solution is that no additional positioning devices are required but just a Bluetooth adapter.

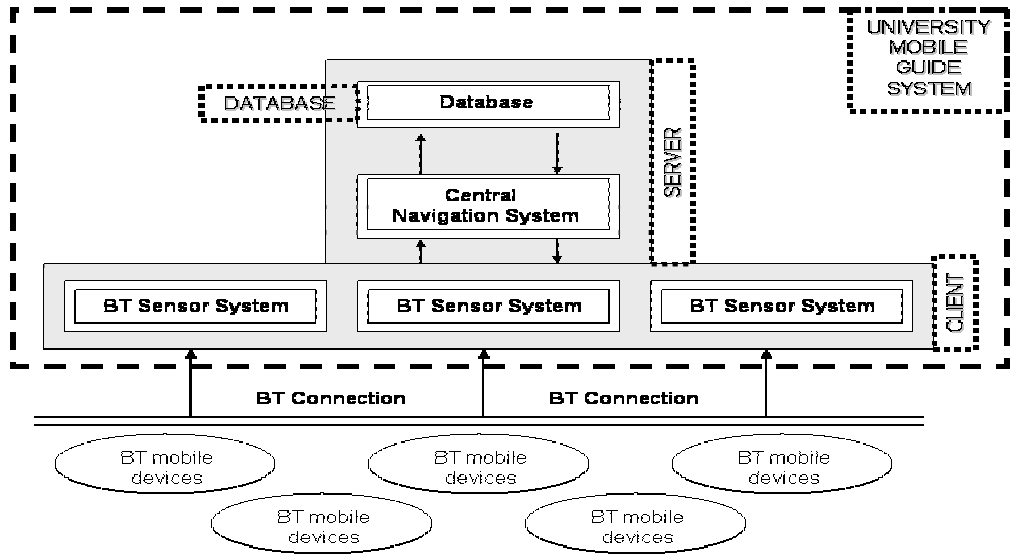

Figure 1. Basic mobile guide system.

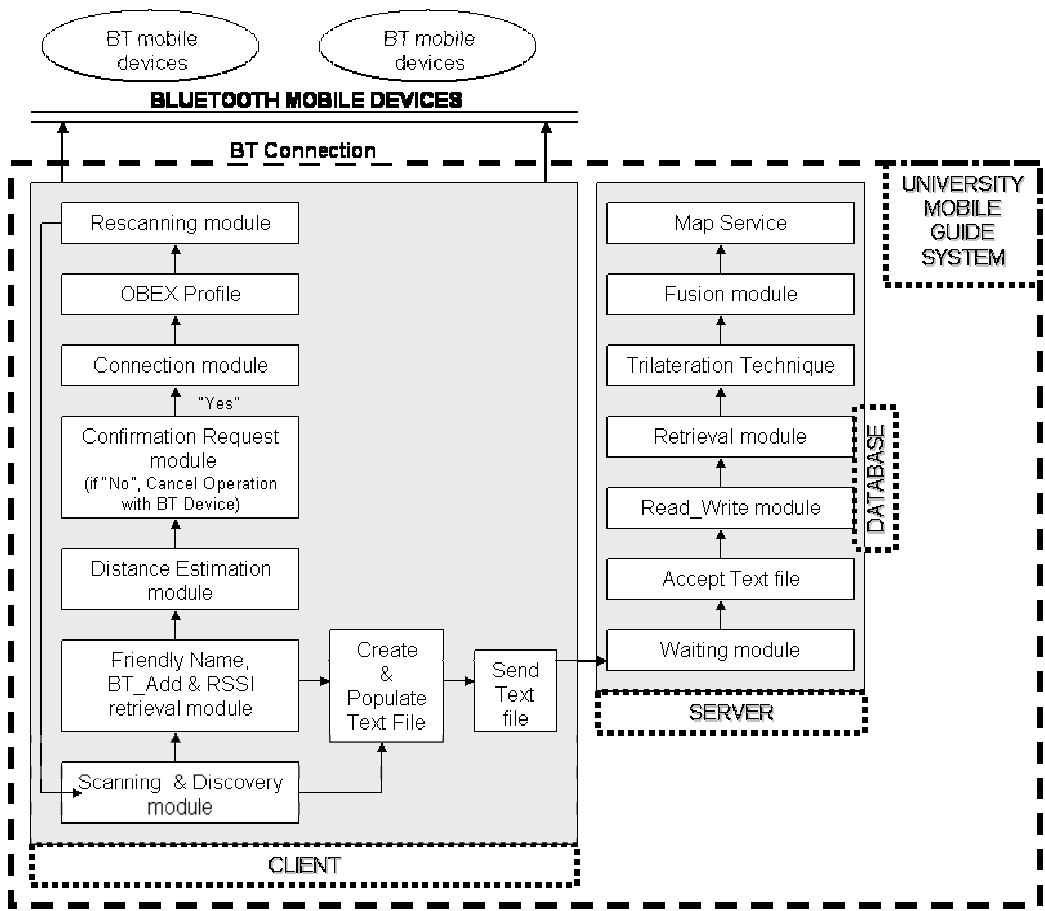

Figure 2. UoMGuide Architecture with location provider

## IV. MOBILE GUIde DESIGN

This UoMGuide architecture, as shown in figure 2, is designed as a location positioning system for accurate position determination of user. In the Scanning & Discovery module, the Bluetooth (BT) Sensor System scans for discoverable BT mobile devices using the BT link controller. When scanning for the BT mobile device, the BT sensor module retrieves all Devices' BT Address and their friendly name using BT Address and Friendly Name Retrieval module. After having scanned and detected the BT mobile devices, a generated text file is populated with the list devices' BT Addresses and is sent to the server. Confirmation Request and Connection module queries for a confirmation to all detected Bluetooth devices if they accept to connect to the Sensor System. The chosen protocol is OBEX and to allow the OBEX connection, RFCOMM is needed as well as the basic lower layer BT protocols. Object PUSH Profile is used to send the predefined customizable text message. The entire process will be executed continuously at a periodic predefined interval of time by the Rescanning module. Once a sensor has connected and sent data, the module accepts all the bitstream that the sensor sends and store all the data in a text file named as that of the sensor IP Address. Read_Write module is responsible to read all the data in the new text file created by the waiting module and to store the data in the database. In the Retrieval module, the numbers of tables present in the database and their respective rows are noted. This information is used to plot the number of BT devices that each sensor has detected on the university map, which is done by the mapping module. This module takes the data from the Retrieval module and plots the data accordingly on the map. Fusion module take the converted distance obtained by the distance estimation module from the different BT sensors and use the trilateration results to combine all these data into a single position location value which will be the location of the BT mobile device. For the development of the UoMGuide system (Sensor and Navigation) Java 2 Development Platform (Java SE Development Kit 6) was used. Microsoft Windows Service Pack 2 is needed for the sensor implementation to be able to detect the BT dongle appropriately. Key libraries for implementation of the BT Sensor System are Avetana Bluetooth, Bluecove Stack, Atinav aveLink Bluetooth, and Java APIs for Bluetooth – javax.Bluetooth [8].

## V. DISCUSSION AND EVALUATION

To deal with the frequent updates from dense and highly dynamic location information, a data storage component (database server) is used. The location server is responsible for tracking all new BT objects visiting the service area. Many of the existing location-based systems and applications do not scale well because they require an expensive infrastructure or a complex location calculation process. Our experiments with the prototype UoMGuide implementation show that our modular architecture is cost effective and scalable. The accuracy and the quality of location information are the requirements that determine the type of services delivered. UoMGuide service ensures interoperability of the system with existing wireless network infrastructure to enable the use and exchange of location information from multiple sources. These factors make the UoMGuide service a success in terms of better usability and the creation of added value for new visitors and students at university. Usability of the system is evaluated as having good user interface, fast response time, low delay and error handling. Personalised and attractive content relevant to the user are delivered at the right moment and location.

## VI. CONCLUSION

There is a growing interest in location-based applications and it can be based on several wireless technologies. We have presented Bluetooth technology as a cost-effective solution for location positioning. Our driving idea was to estimate location with a proximity based positioning approach. We have developed an interoperable UoMGuide LBS application with message delivery to enable the use of spatial information to guide visitors. A future interesting research area is how to integrate readily available infrastructure Wireless LAN access points on campus and Bluetooth as sensors to determine the location of a mobile device for the UoMGuide application.